\documentclass[%
 reprint,
 superscriptaddress,
 amsmath,amssymb,
 aps, 
 prl,
]{revtex4-1}

\usepackage{graphicx}
\usepackage{dcolumn}
\usepackage{bm}
\usepackage{float}
\usepackage{color,soul}
\usepackage{xcolor}
\usepackage{amsmath}
\usepackage{physics}

\begin{document}


\title{Preserving Orbital Order in a Layered Manganite by Ultrafast Hybridized Band-excitation}

\author{L. Shen}
\email{lingjias@slac.stanford.edu}
\affiliation{Stanford Institute for Materials and Energy Science, Stanford University and SLAC National Accelerator Laboratory, Menlo Park, California 94025, USA}
\affiliation{Linac Coherent Light Source, SLAC National Accelerator Laboratory, 2575 Sand Hill Road, Menlo Park, CA 94025, USA}%

\author{S. Mack}
\author{G. Dakovski}
\author{G. Coslovich}
\author{O. Krupin}
\author{M. Hoffmann}
\affiliation{Linac Coherent Light Source, SLAC National Accelerator Laboratory, 2575 Sand Hill Road, Menlo Park, CA 94025, USA}%

\author{S-W. Huang}
\author{Y-D. Chuang}
\affiliation{Advanced Light Source, Lawrence Berkeley National Laboratory, Berkeley, California 94720, USA}

\author{J. A. Johnson}
\affiliation{Swiss Light Source, Paul Scherrer Institut (PSI), 5232 Villigen PSI, Switzerland}


\author{S. Lieu}
\author{S. Zohar}
\author{C. Ford}
\author{M. Kozina}
\author{W. Schlotter}
\author{M. P. Minitti}
\affiliation{Linac Coherent Light Source, SLAC National Accelerator Laboratory, 2575 Sand Hill Road, Menlo Park, CA 94025, USA}%

\author{J. Fujioka}
\affiliation{Department of Applied Physics and Quantum-Phase Electronics Center (QPEC), University of Tokyo, Hongo, Tokyo 113-8656, Japan}

\affiliation{Graduate School of Pure and Applied Science,
University of Tsukuba, Tsukuba, Ibaraki, 305-8573, Japan}

\author{R. Moore}
\author{W-S. Lee}
\affiliation{Stanford Institute for Materials and Energy Science, Stanford University and SLAC National Accelerator Laboratory, Menlo Park, California 94025, USA}



\author{Z. Hussain}
\affiliation{Advanced Light Source, Lawrence Berkeley National Laboratory, Berkeley, California 94720, USA}



\author{Y. Tokura}
\affiliation{Department of Applied Physics and Quantum-Phase Electronics Center (QPEC), University of Tokyo, Hongo, Tokyo 113-8656, Japan}
\affiliation{RIKEN Center for Emergent Matter Science (CEMS), Wako 351-0198, Japan}

\author{P. Littlewood}
\affiliation{James Franck Institute and Department of Physics, University of Chicago, Illinois 60637, USA}%

\author{J. J. Turner}
\email{joshuat@slac.stanford.edu}
\affiliation{Stanford Institute for Materials and Energy Science, Stanford University and SLAC National Accelerator Laboratory, Menlo Park, California 94025, USA}
\affiliation{Linac Coherent Light Source, SLAC National Accelerator Laboratory, 2575 Sand Hill Road, Menlo Park, CA 94025, USA}%

\date{\today}

\begin{abstract}
In the mixed-valence manganites, a near-infrared laser typically melts the orbital and spin order simultaneously, corresponding to the photoinduced $d^{1}d^{0}$ $\xrightarrow{}$ $d^{0}d^{1}$ excitations in the Mott-Hubbard bands of manganese. Here, we use ultrafast methods -- both femtosecond resonant x-ray diffraction and optical reflectivity -- to demonstrate that the orbital response in the layered manganite Nd$_{1-x}$Sr$_{1+x}$MnO$_{4}$ ($\it{x}$\ =\ 2/3) does not follow this scheme. At the photoexcitation saturation fluence, the orbital order is only diminished by a few percent in the transient state. Instead of the typical $d^{1}d^{0}$ $\xrightarrow{}$ $d^{0}d^{1}$ transition, a near-infrared pump in this compound promotes a fundamentally distinct mechanism of charge transfer, the $d^{0}$ $ \xrightarrow{}$ $d^{1}L$, where $\it{L}$ denotes a hole in the oxygen band. This novel finding may pave a new avenue for selectively manipulating specific types of order in complex materials of this class.

\begin{description}
\item[PACS numbers]{78.47.J, 75.47.Lx, 75.10.b, 61.05.C}

\end{description}
\end{abstract}

\maketitle

The correlation effect in quantum solids, which refers to the many-body interactions between charge, orbital, spin, and nuclear lattice, plays an essential role in their electronic structures, determining transport properties \cite{Dagotto,Dagotto2}. One paradigmatic example is the Mott-Hubbard insulator that dominates in the strong Coulomb repulsion (U $\xrightarrow{}$ $\infty$) limit, even when band theory predicts a metallic state \cite{Mott}. Doping a Mott insulator with carriers often results in a rich electronic phase diagram. Notably, these phases and their interactions lie at the heart of understanding some of the most intricate problems in quantum materials including high-temperature superconductivity \cite{keimer-2015-nature}, colossal magnetoresistance \cite{yamada-2019-prl,Dagotto3}, and magnetic and electronic topology \cite{Rau}.

\begin{figure}[b]
	\centering
	\includegraphics[width=0.48\textwidth]{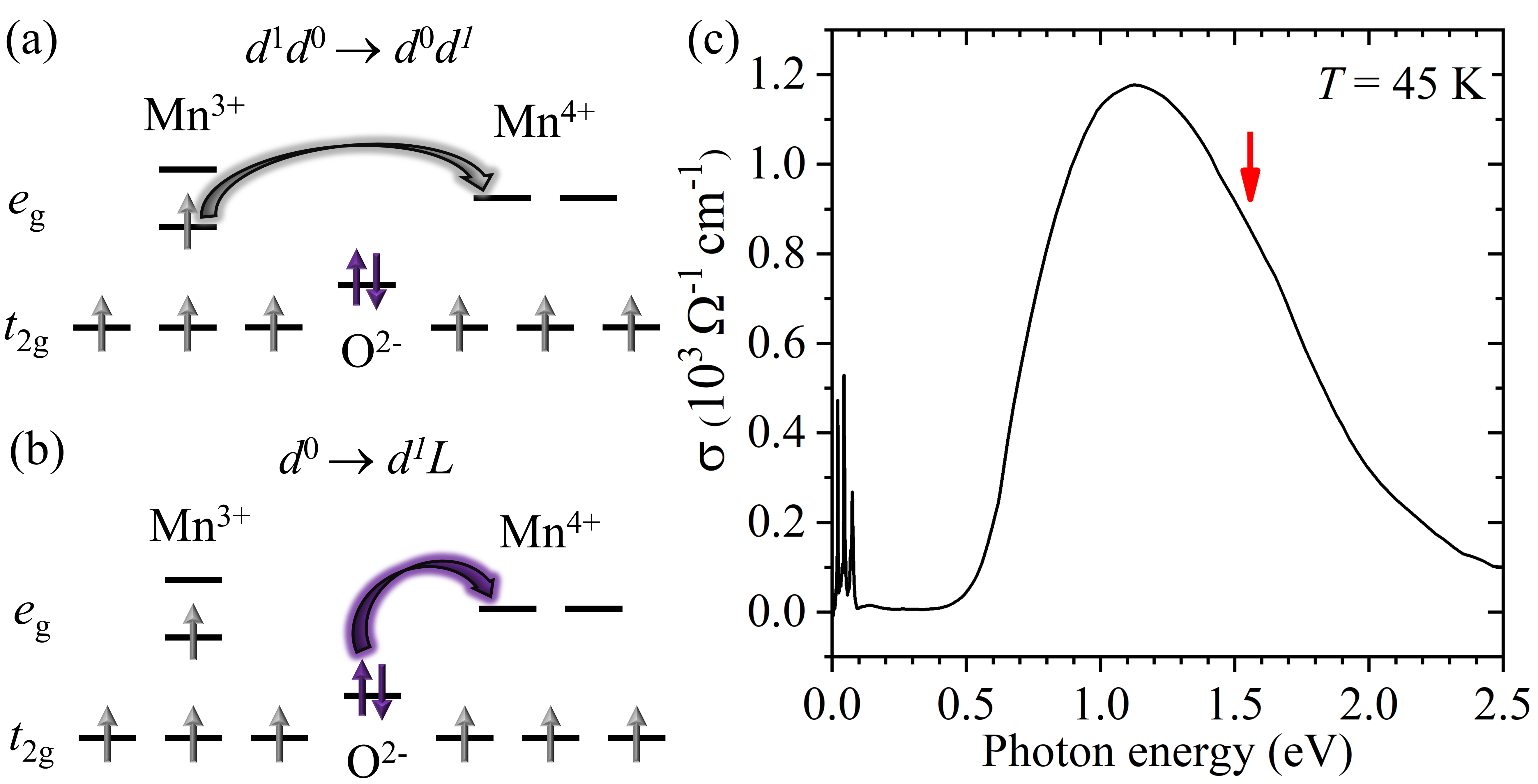}
	\caption{Schematic view of the (a) Mott-Hubbard type $d^{1}d^{0}$ $\xrightarrow{}$ $d^{0}d^{1}$ and (b) charge-transfer type $d^{0}$ $ \xrightarrow{}$ $d^{1}L$ excitations in a mixed-valence manganite. (c) Optical conductivity $\sigma(\omega)$ of NSMO ($\it{x}$\ =\ 2/3) at 45 K. The data is similar to lower temperature data reported in Ref.~\onlinecite{Fujioka}, and was collected in the same fashion. The modes below 0.1 eV come from the phonons and collective density-wave excitations \cite{Fujioka}. The red arrow marks the NIR pump energy (1.55 eV).}
	\label{fig:1}
\end{figure}

In the mixed-valence (Mn$^{3+}$\,/\,Mn$^{4+}$) manganites, electronic correlation is the foundation for the nonthermal manipulation of the order parameter using external stimuli. On the experimental front, such manipulation is of great interest not only because it complements the existing theories in understanding the complex quantum mechanics in relevant systems \cite{Dagotto3}, but also due to the potential for the applications in information technology \cite{Bibes}. Over the past decade, near-infrared (NIR) optical lasers have been broadly utilized to realize the $\it{simultaneous}$ control of the spin and orbital order in manganites on the femto- and pico- second timescales \cite{Matsubara,Li,Beaud,Langner,Zhou}. Theoretically, this phenomenon is governed by the Mott-Hubbard insulating nature of the relevant compounds, i.e. the electronic excitation located at the NIR pump energy is $d^{1}d^{0}$ $\xrightarrow{}$ $d^{0}d^{1}$, for example, between the adjacent Mn$^{3+}$\,/\,Mn$^{4+}$ sites (Fig. \ref{fig:1}a).

According to the Zaanen–Sawatzky–Allen classification scheme, a comprehensive description of the electronic structure near the Fermi surface (FS) of a correlated transition-metal insulator requires not only the Hubbard $U$, but also the charge-transfer energy $\Delta$ \cite{Jan}. In the manganite family, $U$ and $\Delta$ fall in the NIR regime and are often comparable \cite{Arima,Gossling,Mildner}. Thus, the lowest-lying photoinduced electronic excitation which is above the band gap can be of charge-transfer type $d^{0}$ $\xrightarrow{}$ $d^{1}L$, where $\it{d}$ and $\it{L}$ denote the $e_\mathrm{g}$ orbital configuration of the Jahn-Teller (JT) inactive Mn$^{4+}$ cation and the hole in the oxygen anion band, respectively (Fig. \ref{fig:1}b) \cite{Jan}.

The $d^{1}L$ transient state does not contribute to ordered orbital structure because of the two-fold $e_\mathrm{g}$ degeneracy \cite{JTNote}. Consequently, the bulk orbital order of the system, which arises from either the direct Goodenough-Kanamori model or as a
consequence of the cooperative JT distortions of the active Mn$^{3+}$ sites \cite{wilkins-2003-prl}, is not expected to change under this photoexcitation scheme, but this has not been demonstrated. On the other hand, the magnetic order will be affected by the charge-transfer process due to the emergent $e_\mathrm{g}$ electron on the JT inactive Mn$^{4+}$ site (Fig. \ref{fig:1}b). This may lead to an enhancement in the ordered magnetic moment on this site or even collapse of the magnetic structure measured in the equilibrium state. The behavior depends on the degree of electron correlation, but these mechanisms have not been studied on ultrafast timescales in detail.


In this Letter, we test this hypothesis with a rarely explored excitation mechanism, and explore how it will affect the orbital structure. We show that in a mixed-valence manganite, the orbital ordering is robust amidst intense femtosecond laser pulse excitation. This was achieved by employing resonant soft x-ray scattering (RSXS) together with optical spectroscopy to study the ultrafast dynamics in the overdoped layered system Nd$_{1-x}$Sr$_{1+x}$MnO$_{4}$ (NSMO, $\it{x}$\ =\ 2/3) \cite{Kimura}. We optically pumped the magnetic system, demonstrating a strong response within the magnetic state. We then directly studied the orbital modulation in the transient state -- at the photoexcitation saturation fluence -- and found incredible durability. This points to the importance of the $d^{0}$ $ \xrightarrow{}$ $d^{1}L$ charge-transfer photoexcitation process. This  observation will be important for technological applications and draws a fundamental distinction from the $d^{1}d^{0}$ $\xrightarrow{}$ $d^{0}d^{1}$ -type excitation which has been reported in past ultrafast experiments on the manganites (Fig. \ref{fig:1}a $\&$ b) \cite{Matsubara,Li,Beaud,Langner,Zhou}.

A single crystal of NSMO ($\it{x}$\ =\ 2/3) was used in our study and was grown by the floating-zone method \cite{Kimura}. It was polished along the (110) direction. The NIR (1.55 eV) pump, RSXS probe measurements were performed on the SXR instrument at the Linac Coherent Light Source using the RSXS endstation, the FCCD detector, and a recently developed THz optical system \cite{Dakovski-2015-JSR, Bostedt-2016-RMP, Doering-2011-RSI, Turner-2015-JSR}. The pump and probe pulses propagated collinearly to the sample position with $\pi$-polarization. The x-ray beam energy was tuned to the Mn $L_{3}$ edge (641.5 eV) with a bandwidth of 1.3 eV \cite{Tiedtke-2014-OptExpress}, giving an attenuation length of about 70 nm for this geometry. The overall time resolution, calibrated by measuring the response of a sample of gallium phosphide, was about 400 fs \cite{Krupin-2012-OptExp}. In addition, we have performed time-resolved optical reflectivity measurements. A high-power Ti\,:\,sapphire-based laser (1.55 eV) with about 50 fs pulse duration and 120 Hz repetition rate was split and cross-polarized into the pump and probe pulses, which gives a temporal resolution of about 75 fs. The optical penetration depth matches that of the x-rays \footnote{We have carefully calculated the refractive index at the NIR pump energy based on the optical conductivity (Fig. \ref{fig:1}c) to obtain a value of 2.62, giving a penetration depth of 67\,nm. This value was then used for the energy density calculation. The incident pump angles in different geometries have also been taken into account, producing internal angles leading only to small corrections of the energy density value ($<$ 10 $\%$), within the energy density errors.}, in agreement with previous work on related systems \cite{Ehrke}.

\begin{figure}[t]
	\centering
	\includegraphics[width=0.45\textwidth]{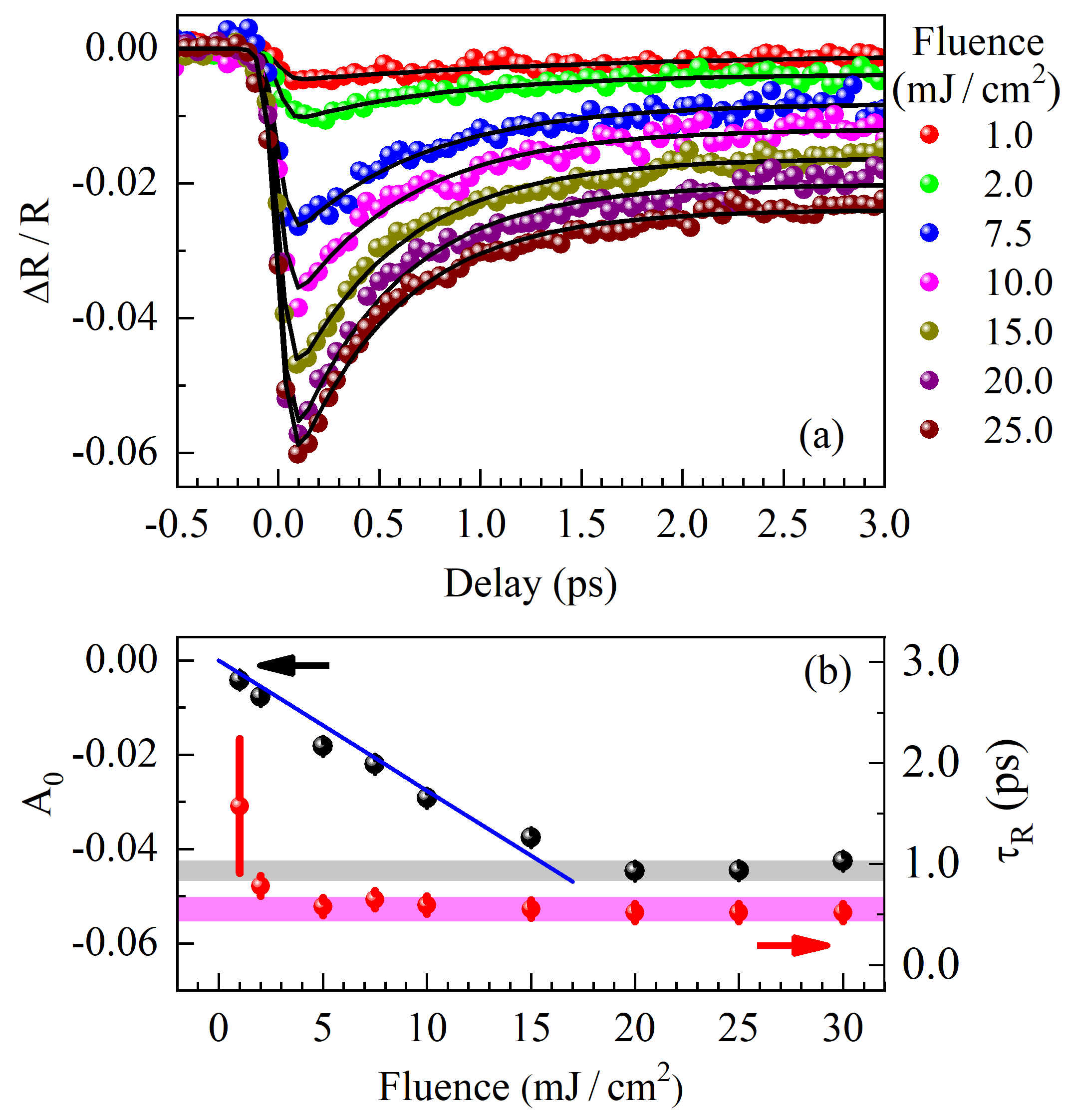}
	\caption{(a) Temporal reflectivity change $\Delta{\mathrm{R}}$\,/\,R of 800\,nm as a function of pump fluence (solid circles). All measurements were performed at 50 K. The solid lines are the numerical fits to Eq. 1. (b) Fluence dependence of the initial decay amplitude $A_{0}$ (black solids, left axis) and relaxation rate ($\mathrm{\tau_{R}}$, red solids, right axis). The blue solid line is a linear fit to the $A_{0}$ data below saturation. The shaded areas are for a guide to the eye.}
	\label{fig:2}
\end{figure}

The NIR optical spectrum in the equilibrium state is shown in Fig. \ref{fig:1}c, where a gapped mode is located between about 0.5 eV and 2.0 eV. This mode is almost temperature independent \cite{Fujioka} and supports the $d^{0}$ $ \xrightarrow{}$ $d^{1}L$ charge-transfer nature of the photoexcitations at these energies \cite{Kovaleva,Tobe}. The ultrafast response of the optical reflectivity $R(t)$ as a function of pump fluence is shown in Fig. \ref{fig:2}a. These curves were fit to a single exponential decay:
\begin{equation}
R\,(t) = \frac{1}{2} \times \left[{\erf}\,\left(\frac{t - t_{0}}{\tau_{0}}\right) + 1\right]
\left[A_{s} + A_{0}e^{-\left({\dfrac{t - t_{0}}{\tau_{R}}}\right)}\right]
\end{equation}
where $\tau_{0}$ and $A_{0}$ are the initial decay rate and amplitude, $\tau_{R}$ is the recovery rate of the fast component, and the constant $A_{s}$ accounts for the long-lived state associated with the nuclear lattice that recovers on a much slower time scale ($\sim$ 400 ps at 21 mJ / cm$^{2}$, data not shown here). All the optical data presented here can be well described by Eq. 1, allowing us to quantitatively determine the electronic response to the NIR pump.

The initial reflectivity drop is always resolution-limited within the errors, i.e.  $\tau_{0}$ = 75 fs, indicating an adiabatic excitation process. The fast recovery process is on the timescale between 0.5 ps and 1.0 ps (Fig. \ref{fig:2}b). This component is often assigned to the electron-electron thermalization \cite{Averitt} that partially recovers the electronic order \cite{Beaud,Langner}. The decay amplitude increases linearly with the fluence until saturating at 16(1) mJ\,/\,cm$^{2}$ (Fig. \ref{fig:2}b). While a NIR pump can also excite free carriers in the system, the linear fluence dependence below the saturation threshold supports the notion that the pump induced reflectivity change at 1.55 eV must mainly come from the above-the-band-gap electronic excitations \cite{Fujioka,Beaud,Esposito,Ehrke}. This is supported by the strongly gapped optical mode (Fig. \ref{fig:1}c). As a result, our optical data have revealed that the electronic system in NSMO ($\it{x}$\ =\ 2/3) has been saturated by the femtosecond NIR pump at $f_\mathrm{sat}$ = 16(1) mJ\,/\,cm$^{2}$, through the ultrafast $d^{0}$ $ \xrightarrow{}$ $d^{1}L$ charge-transfer process (Fig. \ref{fig:1}b). 


\begin{figure}[t]
	\centering
	\includegraphics[width=0.42\textwidth]{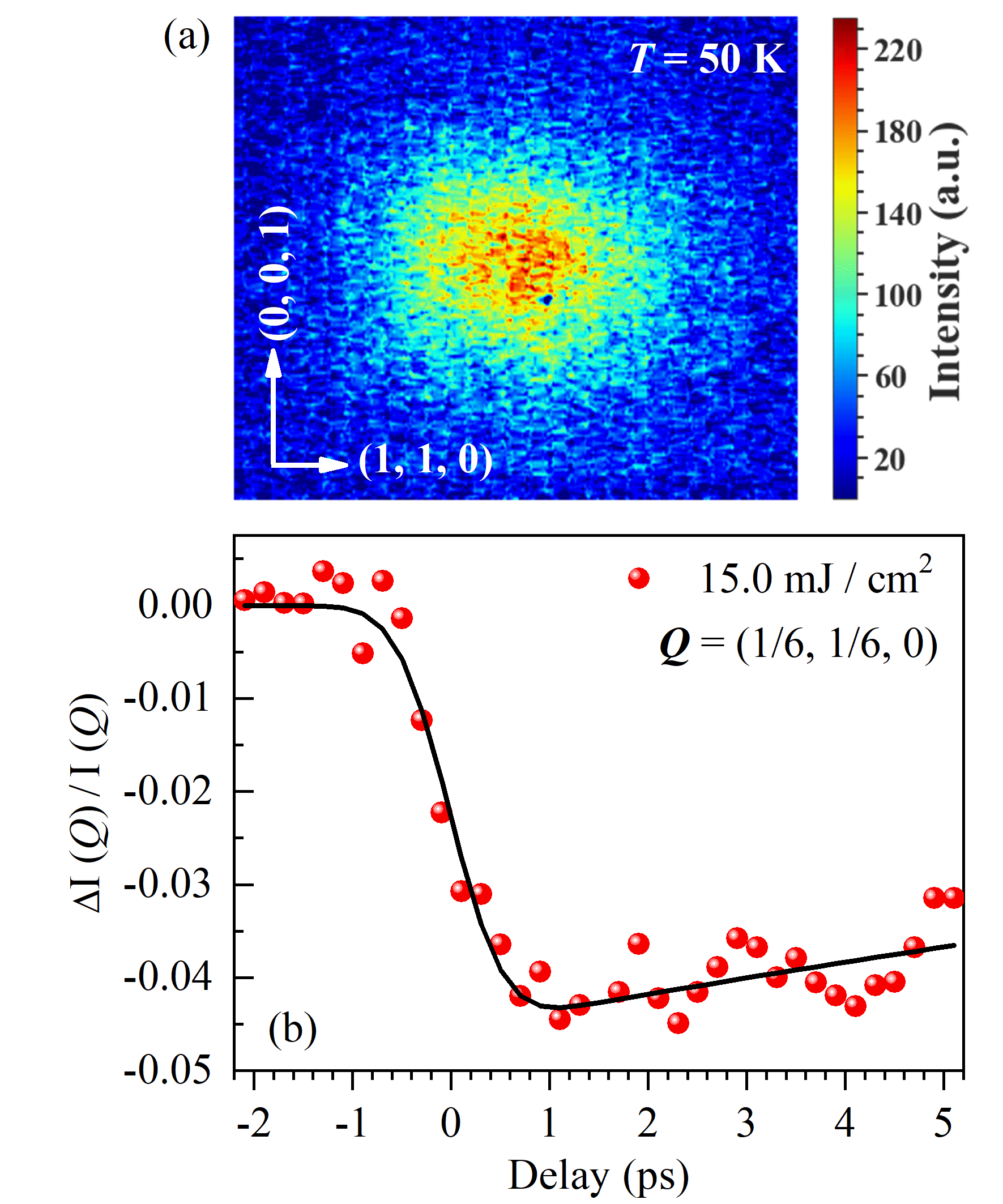}
	\caption{(a) FCCD image (200 $\times$ 120 pixels) of the RSXS diffraction peak from the orbital order around $Q=(1/6, 1/6, 0)$ at 50 K. (b) Temporal evolution of the relative RSXS intensity change [$\Delta{\mathrm{I}\,(Q)}$\,/\,$\mathrm{I}\,(Q)$] at $\it{Q}$ = (1/6, 1/6, 0) recorded at 50 K and different fluences (solid circles). The solid lines are numerical fits to Eq.\,1 by fixing $A_{s}$ to zero.}
	\label{fig:3}
\end{figure}

Typically, the ultrafast dynamics in a manganite under an optical pump are governed by the Mott-Hubbard type $d^{1}d^{0}$ $\xrightarrow{}$ $d^{0}d^{1}$ excitations. This strongly suppresses the underlying orbital order, as has been reported by many authors \cite{Matsubara,Li,Beaud,Langner,Zhou}. We implemented RSXS to investigate if the ultrafast orbital order response matches to a charge-transfer-type excitation.

Figure \ref{fig:3}a shows the RSXS image around $Q= (1/6, 1/6, 0)$ recorded by the FCCD detector at 50 K. A resolution-limited diffraction spot can be clearly visualized. This position was chosen because the modulation vector of the orbital order in materials of this class is predicted to be $\bm{k}$ = ($\delta$, $\delta$, 0), where $\delta$ is determined by the hole concentration: $\delta$ $\approx$ (1\ -\ $\it{x}$)\ /\ 2 \cite{Kimura}. In the NSMO family, an incommensurate superlattice at this position has been observed by electron diffraction \cite{Kimura}, but it has never been directly verified which states contribute to this superstructure. Since resonant scattering at the $L$-edge probes the unoccupied $d$-orbital state, this directly confirms that the modulation is driven by the Mn $d$ orbital states. The correlation length of this orbital state is observed to be 330 \,$\mathrm{\AA}$.

Next the ultrafast dynamics of the orbital order under femtosecond NIR excitation were measured. For 15.0 mJ\,/\,cm$^{2}$, which is slightly below or equal to $f_\mathrm{sat}$ (Fig. \ref{fig:2}b), the RSXS pump-probe curve is shown in Fig. \ref{fig:3}b. This behavior also fits well to Eq.\,1. The initial x-ray scattering intensity decay is resolution-limited, but is given here by the jitter correction of the optical and x-ray pulse arrival time, i.e. $\tau_{0}$ = 400 fs \cite{Beye-2012-APL}. This verifies the adiabatic nature of this photoinduced transition.

For recovery times, the time window is not wide enough to reliably fit the $A_{s}$ term on the longer timescales. Fixing it to zero leads to $\tau_{R}$ = 23(7) ps. This value is smaller than the spin-lattice thermalization, i.e. the energy transfer from the initially excited electron system to the nuclear lattice. This timescale is usually long ($>$ 100 ps) in relevant systems \cite{Ehrke}, but is also considerably longer than the fast electronic recovery process extracted from the optical reflectivity measurements ($\leq$ 1.0 ps, Fig. \ref{fig:2}b). This intermediate recovery process, which has been observed in other non-optimally-doped manganites \cite{Langner}, is related to the transient photoinduced phase separation that leads to an inhomogenous recovery of the electronic order. 

Most importantly, we have found that the orbital order is surprisingly robust against this photoexcitation. Although the electronic system is saturated by the NIR pump at 15.0 mJ\,/\,cm$^{2}$ (Fig. \ref{fig:2}b), the RSXS intensity scattered by the orbital order is only suppressed by 4.5(2)\,$\%$ in the transient state. Because we have carefully calibrated the energy density for the different geometries and experiments, this character sharply contrasts the observations in other manganites \footnotemark[\value{footnote}]. In previous experiments on similar systems, the same NIR excitation promotes the Mott-Hubbard type $d^{1}d^{0}$ $\xrightarrow{}$ $d^{0}d^{1}$ transitions (Fig. \ref{fig:1}a) and therefore greatly suppresses the underlying orbital order \cite{Beaud}.

\begin{figure}[b]
	\centering
	\includegraphics[width=0.49\textwidth]{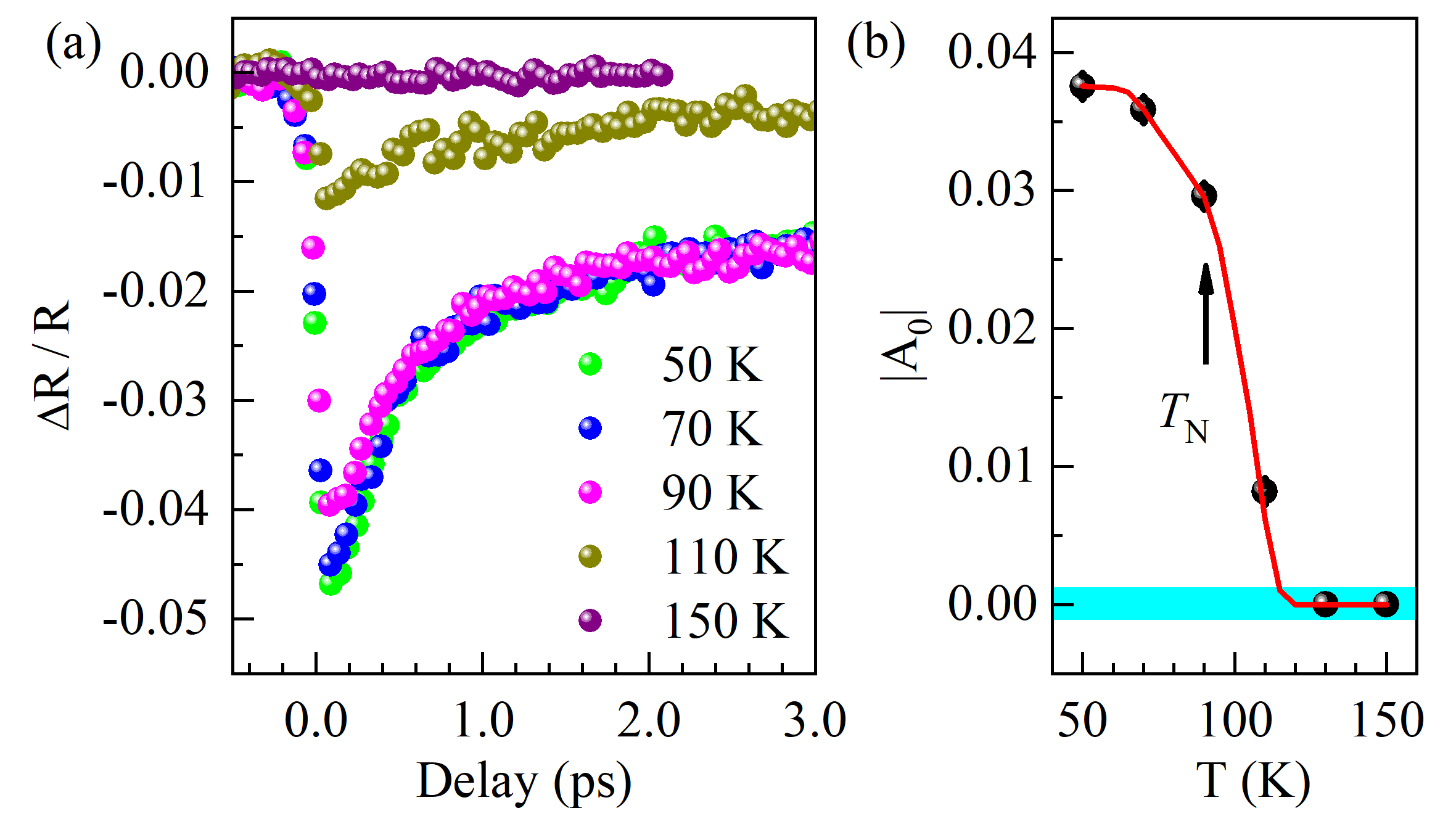}
	\caption{(a) Temporal evolution of the relative optical reflectivity change ($\Delta{\mathrm{R}}$\,/\,R) as a function of temperature. All the curves were recorded at the pump fluence 15 mJ\,/\,cm$^{2}$. (b) Temperature dependence of the absolute value of the initial decay amplitude ($A_{0}$) extracted from Eq. 2 (solid circles). The red line and shaded area are guides for the eye.}
	\label{fig:4}
\end{figure}

Due to the comparable $U$ and $\Delta$ values in mixed-valence manganites, the above-the-band-gap excitation near the FS can to be dominated by either the $d^{1}d^{0}$ $\xrightarrow{}$ $d^{0}d^{1}$ transition, the $d^{0}$ $ \xrightarrow{}$ $d^{1}L$, or both \cite{Gossling,Mildner}. Our RSXS observations rule out a meaningful contribution from the former in NSMO ($\it{x}$\ =\ 2/3), since the orbital order intensity loss of only 4.5(2)\,$\%$ in the transient state is measured close to $f_\mathrm{sat}$. The robustness of the orbital order therefore strongly indicates that the latter charge-transfer type scheme is responsible for the NIR photoexcitation process in this compound (Fig. \ref{fig:1}b). This is consistent with the nearly temperature independent optical spectrum reported at these energies \cite{Fujioka}. 

To obtain more insight about the nature of the large optical spectral weight at 1.55 eV in Fig. \ref{fig:1}c, we measured the ultrafast dynamics of the optical reflectivity at 15 mJ\,/\,cm$^{2}$ as a function of temperature (Fig. \ref{fig:4}). The decay amplitude $A_{0}$ changes smoothly with temperature below about 90 K, which is at the N\'eel temperature ($T_\mathrm{N}$) of the charge-exchange-like antiferromagnetic (AFM) order in this compound \cite{Ulbrich2}. $A_{0}$ is suppressed upon heating above $T_\mathrm{N}$ and falls below the noise level around 110 K (Fig. \ref{fig:4}b).

The electronic excitation process triggered by the NIR pump is activated by the increased probability amplitude for the hopping matrix element -- in the magnetic state -- but originating from the hybridized wavefunction. Intuitively, this character does not fit the $d^{0}$ $\xrightarrow{}$ $d^{1}L$ scenario depicted in Fig.\,\ref{fig:1}b because it does not require magnetic order. However, similar behaviour is expected if the $e_\mathrm{g}$ band formed by the Mn$^{3+}$ cations is involved. For example, the $d^{1}d^{0}$ $\xrightarrow{}$ $d^{0}d^{1}$ process in Fig. \ref{fig:1}a belongs to this category; it is energetically favored if the adjacent Mn$^{3+}$\,/\,Mn$^{4+}$ spins are parallel with each other, a configuration that is only fulfilled in the magnetically ordered region of the relevant manganites \cite{Ehrke,Beaud,Zhou}.

Since the orbital response implies the $d^{1}d^{0}$ $\xrightarrow{}$ $d^{0}d^{1}$ contribution is very weak in NSMO ($\it{x}$\ =\ 2/3), we propose that the $d^{0}$ $\xrightarrow{}$ $d^{1}L$ process is between the hybridized Mn (3$\it{d}$) and O (2$\it{p}$) bands near the FS \cite{Fujioka}. The $\it{dp}$ hybridization occurs when $U$ and $\Delta$ are similar in strength \cite{Gossling,Mildner}. In this modified scenario, the O$^{2-}$ 2$\it{p}$ electrons are spin polarized by the Mn$^{3+}$ $e_\mathrm{g}$ electrons in the hybridized $\it{dp}$ band. Accordingly, the photoinduced charge-transfer is allowed along the ferromagnetic zigzag spin chains along the (1, 1, 0) direction, which only exists in the AFM state of NSMO ($\it{x}$\ =\ 2/3) \cite{Ulbrich}.

The results presented here establish selective `band-based' excitation in systems with multiple types of order on the same energy scale. Typically, ultrafast intense optical illumination in the mixed-valence manganites drives the simultaneous change of all order parameters, e.g. spin, orbital and charge \cite{Zhou,Ehrke,Beaud}. Selective manipulation of a single order parameter, while keeping the others unperturbed, is of great importance to the fabrication of multistate logic or memory devices \cite{Spaldin}. This is known to be challenging because of the strong electron correlation effect and has not been demonstrated on the femtosecond timescale in any material of this class. 
The realization of such manipulation of the magnetic order is highlighted by the robust orbital order while the electronic system is fully photoexcited.

Furthermore, the $d^{0}$ $\xrightarrow{}$ $d^{1}L$ scenario may also be partially responsible for the ultrafast orbital dynamics in the other layered manganite La$_{1-x}$Sr$_{1+x}$MnO$_{4}$ ($\it{x}$\ =\ 1/2) \cite{Ehrke}. In this compound, a femtosecond NIR pump melts about 25\,$\%$ of the orbital order reflection probed by RSXS at the saturation fluence, whereas the AFM order is completely suppressed. This finite RSXS intensity was explained by the residual JT distortion after the photoexcitation \cite{Ehrke}. A recent femtosecond NIR pump / hard x-ray probe work on Pr$_{1-x}$Ca$_{x}$MnO$_{3}$ ($\it{x}$\ =\ 1/2), where the $d^{1}d^{0}$ $\xrightarrow{}$ $d^{0}d^{1}$ physics is clearly dominant, directly revealed that the JT distortion disappears with the orbital and charge order when the electronic system is fully excited \cite{Beaud} and therefore does not fit the model proposed in Ref.~\onlinecite{Ehrke}. Alternatively, the residual orbital order in La$_{1-x}$Sr$_{1+x}$MnO$_{4}$ ($\it{x}$\ =\ 1/2) could indicate that about 75 $\%$ of the electronic excitations involve the $d^{0}$ $\xrightarrow{}$ $d^{1}L$ charge-transfer.

In conclusion, we have investigated the effect of NIR excitation (1.55 eV) on the orbital order in the layered manganite NSMO ($\it{x}$\ =\ 2/3) through ultrafast techniques. Conducting a systematic ultrafast optical reflectivity study, we found the fluence and temperature dependence of the optical reflectivity change. By using RSXS at an x-ray free-electron laser source, we combined this finding with the ability to directly monitor the femtosecond orbital response in the transient photoexcited state. We found that the RSXS intensity arising from the orbital order is diminished by no more than 5 $\%$ percent while the electronic system is fully saturated. These results strongly suggest that the photoexcitation mechanism in this compound is $d^{0}$ $\xrightarrow{}$ $d^{1}L$, instead of the commonly assumed $d^{1}d^{0}$ $\xrightarrow{}$ $d^{0}d^{1}$. In addition, we show that the correct description of the photoexcitation process requires the $dp$ hybridization near the FS in the magnetically ordered state. Taking advantage of the $d^{0}$ $\xrightarrow{}$ $d^{1}L$ charge-transfer mechanism holds great potential for selectively manipulating the electronic order. Furthermore, this shows that other excitation mechanisms at play may be more closely related to that seen in the cuprates, such as the high-energy scale physics associated with Mott-like excitations \cite{giannetti-natcomm-2011}. The results here point to future work based on band-excitation to unravel the different excitation mechanisms in systems with multiple types or order, such as other quantum materials that display the nearly degenerate energy scales as observed in the manganites.


\begin{acknowledgments}


We acknowledge useful discussions with R. Schoenlein, A. Sakdinawat, S. Johnson, and U. Staub.
This work is supported by the U. S. Department of Energy, Office of Science, Basic Energy Sciences, Materials Sciences and Engineering Division, under Contract DE-AC02-76SF00515. The use of the Linac Coherent Light Source (LCLS), SLAC National Accelerator Laboratory, is also supported under the same contract. J. J. Turner acknowledges support from the U.S. DOE, Office of Science, Basic Energy Sciences through the Early Career Research Program. Y.-D. C. acknowledges the support from the Advanced Light Source, a DOE Office of Science User Facility under contract no. DE-AC02-05CH11231.
\end{acknowledgments}

%


\begin{thebibliography}{38}%
\makeatletter
\providecommand \@ifxundefined [1]{%
 \@ifx{#1\undefined}
}%
\providecommand \@ifnum [1]{%
 \ifnum #1\expandafter \@firstoftwo
 \else \expandafter \@secondoftwo
 \fi
}%
\providecommand \@ifx [1]{%
 \ifx #1\expandafter \@firstoftwo
 \else \expandafter \@secondoftwo
 \fi
}%
\providecommand \natexlab [1]{#1}%
\providecommand \enquote  [1]{``#1''}%
\providecommand \bibnamefont  [1]{#1}%
\providecommand \bibfnamefont [1]{#1}%
\providecommand \citenamefont [1]{#1}%
\providecommand \href@noop [0]{\@secondoftwo}%
\providecommand \href [0]{\begingroup \@sanitize@url \@href}%
\providecommand \@href[1]{\@@startlink{#1}\@@href}%
\providecommand \@@href[1]{\endgroup#1\@@endlink}%
\providecommand \@sanitize@url [0]{\catcode `\\12\catcode `\$12\catcode
  `\&12\catcode `\#12\catcode `\^12\catcode `\_12\catcode `\%12\relax}%
\providecommand \@@startlink[1]{}%
\providecommand \@@endlink[0]{}%
\providecommand \url  [0]{\begingroup\@sanitize@url \@url }%
\providecommand \@url [1]{\endgroup\@href {#1}{\urlprefix }}%
\providecommand \urlprefix  [0]{URL }%
\providecommand \Eprint [0]{\href }%
\providecommand \doibase [0]{http://dx.doi.org/}%
\providecommand \selectlanguage [0]{\@gobble}%
\providecommand \bibinfo  [0]{\@secondoftwo}%
\providecommand \bibfield  [0]{\@secondoftwo}%
\providecommand \translation [1]{[#1]}%
\providecommand \BibitemOpen [0]{}%
\providecommand \bibitemStop [0]{}%
\providecommand \bibitemNoStop [0]{.\EOS\space}%
\providecommand \EOS [0]{\spacefactor3000\relax}%
\providecommand \BibitemShut  [1]{\csname bibitem#1\endcsname}%
\let\auto@bib@innerbib\@empty
\bibitem [{\citenamefont {Dagotto}(1994)}]{Dagotto}%
  \BibitemOpen
  \bibfield  {author} {\bibinfo {author} {\bibfnamefont {E.}~\bibnamefont
  {Dagotto}},\ }\href {\doibase 10.1103/RevModPhys.66.763} {\bibfield
  {journal} {\bibinfo  {journal} {Rev. Mod. Phys.}\ }\textbf {\bibinfo {volume}
  {66}},\ \bibinfo {pages} {763} (\bibinfo {year} {1994})}\BibitemShut
  {NoStop}%
\bibitem [{\citenamefont {Dagotto}(2005)}]{Dagotto2}%
  \BibitemOpen
  \bibfield  {author} {\bibinfo {author} {\bibfnamefont {E.}~\bibnamefont
  {Dagotto}},\ }\href {\doibase 10.1126/science.1107559} {\bibfield  {journal}
  {\bibinfo  {journal} {Science}\ }\textbf {\bibinfo {volume} {309}},\ \bibinfo
  {pages} {257} (\bibinfo {year} {2005})}\BibitemShut {NoStop}%
\bibitem [{\citenamefont {Mott}(1949)}]{Mott}%
  \BibitemOpen
  \bibfield  {author} {\bibinfo {author} {\bibfnamefont {N.~F.}\ \bibnamefont
  {Mott}},\ }\href {http://stacks.iop.org/0370-1298/62/i=7/a=303} {\bibfield
  {journal} {\bibinfo  {journal} {Proceedings of the Physical Society. Section
  A}\ }\textbf {\bibinfo {volume} {62}},\ \bibinfo {pages} {416} (\bibinfo
  {year} {1949})}\BibitemShut {NoStop}%
\bibitem [{\citenamefont {Keimer}\ \emph {et~al.}(2015)\citenamefont {Keimer},
  \citenamefont {Kivelson}, \citenamefont {Norman}, \citenamefont {Uchida},\
  and\ \citenamefont {Zaanen}}]{keimer-2015-nature}%
  \BibitemOpen
  \bibfield  {author} {\bibinfo {author} {\bibfnamefont {B.}~\bibnamefont
  {Keimer}}, \bibinfo {author} {\bibfnamefont {S.~A.}\ \bibnamefont
  {Kivelson}}, \bibinfo {author} {\bibfnamefont {M.~R.}\ \bibnamefont
  {Norman}}, \bibinfo {author} {\bibfnamefont {S.}~\bibnamefont {Uchida}}, \
  and\ \bibinfo {author} {\bibfnamefont {J.}~\bibnamefont {Zaanen}},\ }\href
  {https://doi.org/10.1038/nature14165} {\bibfield  {journal} {\bibinfo
  {journal} {Nature}\ }\textbf {\bibinfo {volume} {518}},\ \bibinfo {pages}
  {179 EP } (\bibinfo {year} {2015})}\BibitemShut {NoStop}%
\bibitem [{\citenamefont {Yamada}\ \emph {et~al.}(2019)\citenamefont {Yamada},
  \citenamefont {Abe}, \citenamefont {Sagayama}, \citenamefont {Ogawa},
  \citenamefont {Yamagami},\ and\ \citenamefont {Arima}}]{yamada-2019-prl}%
  \BibitemOpen
  \bibfield  {author} {\bibinfo {author} {\bibfnamefont {S.}~\bibnamefont
  {Yamada}}, \bibinfo {author} {\bibfnamefont {N.}~\bibnamefont {Abe}},
  \bibinfo {author} {\bibfnamefont {H.}~\bibnamefont {Sagayama}}, \bibinfo
  {author} {\bibfnamefont {K.}~\bibnamefont {Ogawa}}, \bibinfo {author}
  {\bibfnamefont {T.}~\bibnamefont {Yamagami}}, \ and\ \bibinfo {author}
  {\bibfnamefont {T.}~\bibnamefont {Arima}},\ }\href {\doibase
  10.1103/PhysRevLett.123.126602} {\bibfield  {journal} {\bibinfo  {journal}
  {Phys. Rev. Lett.}\ }\textbf {\bibinfo {volume} {123}},\ \bibinfo {pages}
  {126602} (\bibinfo {year} {2019})}\BibitemShut {NoStop}%
\bibitem [{\citenamefont {Dagotto}\ \emph {et~al.}(2001)\citenamefont
  {Dagotto}, \citenamefont {Hotta},\ and\ \citenamefont {Moreo}}]{Dagotto3}%
  \BibitemOpen
  \bibfield  {author} {\bibinfo {author} {\bibfnamefont {E.}~\bibnamefont
  {Dagotto}}, \bibinfo {author} {\bibfnamefont {T.}~\bibnamefont {Hotta}}, \
  and\ \bibinfo {author} {\bibfnamefont {A.}~\bibnamefont {Moreo}},\ }\href
  {\doibase https://doi.org/10.1016/S0370-1573(00)00121-6} {\bibfield
  {journal} {\bibinfo  {journal} {Physics Reports}\ }\textbf {\bibinfo {volume}
  {344}},\ \bibinfo {pages} {1 } (\bibinfo {year} {2001})}\BibitemShut
  {NoStop}%
\bibitem [{\citenamefont {Rau}\ \emph {et~al.}(2016)\citenamefont {Rau},
  \citenamefont {Lee},\ and\ \citenamefont {Kee}}]{Rau}%
  \BibitemOpen
  \bibfield  {author} {\bibinfo {author} {\bibfnamefont {J.~G.}\ \bibnamefont
  {Rau}}, \bibinfo {author} {\bibfnamefont {E.~K.-H.}\ \bibnamefont {Lee}}, \
  and\ \bibinfo {author} {\bibfnamefont {H.-Y.}\ \bibnamefont {Kee}},\ }\href
  {\doibase 10.1146/annurev-conmatphys-031115-011319} {\bibfield  {journal}
  {\bibinfo  {journal} {Annual Review of Condensed Matter Physics}\ }\textbf
  {\bibinfo {volume} {7}},\ \bibinfo {pages} {195} (\bibinfo {year}
  {2016})}\BibitemShut {NoStop}%
\bibitem [{\citenamefont {{Bibes}}\ and\ \citenamefont
  {{Barthelemy}}(2007)}]{Bibes}%
  \BibitemOpen
  \bibfield  {author} {\bibinfo {author} {\bibfnamefont {M.}~\bibnamefont
  {{Bibes}}}\ and\ \bibinfo {author} {\bibfnamefont {A.}~\bibnamefont
  {{Barthelemy}}},\ }\href {\doibase 10.1109/TED.2007.894366} {\bibfield
  {journal} {\bibinfo  {journal} {IEEE Transactions on Electron Devices}\
  }\textbf {\bibinfo {volume} {54}},\ \bibinfo {pages} {1003} (\bibinfo {year}
  {2007})}\BibitemShut {NoStop}%
\bibitem [{\citenamefont {Matsubara}\ \emph {et~al.}(2007)\citenamefont
  {Matsubara}, \citenamefont {Okimoto}, \citenamefont {Ogasawara},
  \citenamefont {Tomioka}, \citenamefont {Okamoto},\ and\ \citenamefont
  {Tokura}}]{Matsubara}%
  \BibitemOpen
  \bibfield  {author} {\bibinfo {author} {\bibfnamefont {M.}~\bibnamefont
  {Matsubara}}, \bibinfo {author} {\bibfnamefont {Y.}~\bibnamefont {Okimoto}},
  \bibinfo {author} {\bibfnamefont {T.}~\bibnamefont {Ogasawara}}, \bibinfo
  {author} {\bibfnamefont {Y.}~\bibnamefont {Tomioka}}, \bibinfo {author}
  {\bibfnamefont {H.}~\bibnamefont {Okamoto}}, \ and\ \bibinfo {author}
  {\bibfnamefont {Y.}~\bibnamefont {Tokura}},\ }\href {\doibase
  10.1103/PhysRevLett.99.207401} {\bibfield  {journal} {\bibinfo  {journal}
  {Phys. Rev. Lett.}\ }\textbf {\bibinfo {volume} {99}},\ \bibinfo {pages}
  {207401} (\bibinfo {year} {2007})}\BibitemShut {NoStop}%
\bibitem [{\citenamefont {Li}\ \emph {et~al.}(2013)\citenamefont {Li},
  \citenamefont {Patz}, \citenamefont {Mouchliadis}, \citenamefont {Yan},
  \citenamefont {Lograsso}, \citenamefont {Perakis},\ and\ \citenamefont
  {Wang}}]{Li}%
  \BibitemOpen
  \bibfield  {author} {\bibinfo {author} {\bibfnamefont {T.}~\bibnamefont
  {Li}}, \bibinfo {author} {\bibfnamefont {A.}~\bibnamefont {Patz}}, \bibinfo
  {author} {\bibfnamefont {L.}~\bibnamefont {Mouchliadis}}, \bibinfo {author}
  {\bibfnamefont {J.}~\bibnamefont {Yan}}, \bibinfo {author} {\bibfnamefont
  {T.~A.}\ \bibnamefont {Lograsso}}, \bibinfo {author} {\bibfnamefont {I.~E.}\
  \bibnamefont {Perakis}}, \ and\ \bibinfo {author} {\bibfnamefont
  {J.}~\bibnamefont {Wang}},\ }\href {https://doi.org/10.1038/nature11934}
  {\bibfield  {journal} {\bibinfo  {journal} {Nature}\ }\textbf {\bibinfo
  {volume} {496}},\ \bibinfo {pages} {69 EP } (\bibinfo {year}
  {2013})}\BibitemShut {NoStop}%
\bibitem [{\citenamefont {Beaud}\ \emph {et~al.}(2014)\citenamefont {Beaud},
  \citenamefont {Caviezel}, \citenamefont {Mariager}, \citenamefont {Rettig},
  \citenamefont {Ingold}, \citenamefont {Dornes}, \citenamefont {Huang},
  \citenamefont {Johnson}, \citenamefont {Radovic}, \citenamefont {Huber},
  \citenamefont {Kubacka}, \citenamefont {Ferrer}, \citenamefont {Lemke},
  \citenamefont {Chollet}, \citenamefont {Zhu}, \citenamefont {Glownia},
  \citenamefont {Sikorski}, \citenamefont {Robert}, \citenamefont {Wadati},
  \citenamefont {Nakamura}, \citenamefont {Kawasaki}, \citenamefont {Tokura},
  \citenamefont {Johnson},\ and\ \citenamefont {Staub}}]{Beaud}%
  \BibitemOpen
  \bibfield  {author} {\bibinfo {author} {\bibfnamefont {P.}~\bibnamefont
  {Beaud}}, \bibinfo {author} {\bibfnamefont {A.}~\bibnamefont {Caviezel}},
  \bibinfo {author} {\bibfnamefont {S.~O.}\ \bibnamefont {Mariager}}, \bibinfo
  {author} {\bibfnamefont {L.}~\bibnamefont {Rettig}}, \bibinfo {author}
  {\bibfnamefont {G.}~\bibnamefont {Ingold}}, \bibinfo {author} {\bibfnamefont
  {C.}~\bibnamefont {Dornes}}, \bibinfo {author} {\bibfnamefont {S.-W.}\
  \bibnamefont {Huang}}, \bibinfo {author} {\bibfnamefont {J.~A.}\ \bibnamefont
  {Johnson}}, \bibinfo {author} {\bibfnamefont {M.}~\bibnamefont {Radovic}},
  \bibinfo {author} {\bibfnamefont {T.}~\bibnamefont {Huber}}, \bibinfo
  {author} {\bibfnamefont {T.}~\bibnamefont {Kubacka}}, \bibinfo {author}
  {\bibfnamefont {A.}~\bibnamefont {Ferrer}}, \bibinfo {author} {\bibfnamefont
  {H.~T.}\ \bibnamefont {Lemke}}, \bibinfo {author} {\bibfnamefont
  {M.}~\bibnamefont {Chollet}}, \bibinfo {author} {\bibfnamefont
  {D.}~\bibnamefont {Zhu}}, \bibinfo {author} {\bibfnamefont {J.~M.}\
  \bibnamefont {Glownia}}, \bibinfo {author} {\bibfnamefont {M.}~\bibnamefont
  {Sikorski}}, \bibinfo {author} {\bibfnamefont {A.}~\bibnamefont {Robert}},
  \bibinfo {author} {\bibfnamefont {H.}~\bibnamefont {Wadati}}, \bibinfo
  {author} {\bibfnamefont {M.}~\bibnamefont {Nakamura}}, \bibinfo {author}
  {\bibfnamefont {M.}~\bibnamefont {Kawasaki}}, \bibinfo {author}
  {\bibfnamefont {Y.}~\bibnamefont {Tokura}}, \bibinfo {author} {\bibfnamefont
  {S.~L.}\ \bibnamefont {Johnson}}, \ and\ \bibinfo {author} {\bibfnamefont
  {U.}~\bibnamefont {Staub}},\ }\href {http://dx.doi.org/10.1038/nmat4046}
  {\bibfield  {journal} {\bibinfo  {journal} {Nature Materials}\ }\textbf
  {\bibinfo {volume} {13}},\ \bibinfo {pages} {923} (\bibinfo {year}
  {2014})}\BibitemShut {NoStop}%
\bibitem [{\citenamefont {Langner}\ \emph {et~al.}(2015)\citenamefont
  {Langner}, \citenamefont {Zhou}, \citenamefont {Coslovich}, \citenamefont
  {Chuang}, \citenamefont {Zhu}, \citenamefont {Robinson}, \citenamefont
  {Schlotter}, \citenamefont {Turner}, \citenamefont {Minitti}, \citenamefont
  {Moore}, \citenamefont {Lee}, \citenamefont {Lu}, \citenamefont {Doering},
  \citenamefont {Denes}, \citenamefont {Tomioka}, \citenamefont {Tokura},
  \citenamefont {Kaindl},\ and\ \citenamefont {Schoenlein}}]{Langner}%
  \BibitemOpen
  \bibfield  {author} {\bibinfo {author} {\bibfnamefont {M.~C.}\ \bibnamefont
  {Langner}}, \bibinfo {author} {\bibfnamefont {S.}~\bibnamefont {Zhou}},
  \bibinfo {author} {\bibfnamefont {G.}~\bibnamefont {Coslovich}}, \bibinfo
  {author} {\bibfnamefont {Y.-D.}\ \bibnamefont {Chuang}}, \bibinfo {author}
  {\bibfnamefont {Y.}~\bibnamefont {Zhu}}, \bibinfo {author} {\bibfnamefont
  {J.~S.}\ \bibnamefont {Robinson}}, \bibinfo {author} {\bibfnamefont {W.~F.}\
  \bibnamefont {Schlotter}}, \bibinfo {author} {\bibfnamefont {J.~J.}\
  \bibnamefont {Turner}}, \bibinfo {author} {\bibfnamefont {M.~P.}\
  \bibnamefont {Minitti}}, \bibinfo {author} {\bibfnamefont {R.~G.}\
  \bibnamefont {Moore}}, \bibinfo {author} {\bibfnamefont {W.~S.}\ \bibnamefont
  {Lee}}, \bibinfo {author} {\bibfnamefont {D.~H.}\ \bibnamefont {Lu}},
  \bibinfo {author} {\bibfnamefont {D.}~\bibnamefont {Doering}}, \bibinfo
  {author} {\bibfnamefont {P.}~\bibnamefont {Denes}}, \bibinfo {author}
  {\bibfnamefont {Y.}~\bibnamefont {Tomioka}}, \bibinfo {author} {\bibfnamefont
  {Y.}~\bibnamefont {Tokura}}, \bibinfo {author} {\bibfnamefont {R.~A.}\
  \bibnamefont {Kaindl}}, \ and\ \bibinfo {author} {\bibfnamefont {R.~W.}\
  \bibnamefont {Schoenlein}},\ }\href {\doibase 10.1103/PhysRevB.92.155148}
  {\bibfield  {journal} {\bibinfo  {journal} {Phys. Rev. B}\ }\textbf {\bibinfo
  {volume} {92}},\ \bibinfo {pages} {155148} (\bibinfo {year}
  {2015})}\BibitemShut {NoStop}%
\bibitem [{\citenamefont {Zhou}\ \emph {et~al.}(2014)\citenamefont {Zhou},
  \citenamefont {Langner}, \citenamefont {Zhu}, \citenamefont {Chuang},
  \citenamefont {Rini}, \citenamefont {Glover}, \citenamefont {Hertlein},
  \citenamefont {Gonzalez}, \citenamefont {Tahir}, \citenamefont {Tomioka},
  \citenamefont {Tokura}, \citenamefont {Hussain},\ and\ \citenamefont
  {Schoenlein}}]{Zhou}%
  \BibitemOpen
  \bibfield  {author} {\bibinfo {author} {\bibfnamefont {S.~Y.}\ \bibnamefont
  {Zhou}}, \bibinfo {author} {\bibfnamefont {M.~C.}\ \bibnamefont {Langner}},
  \bibinfo {author} {\bibfnamefont {Y.}~\bibnamefont {Zhu}}, \bibinfo {author}
  {\bibfnamefont {Y.-D.}\ \bibnamefont {Chuang}}, \bibinfo {author}
  {\bibfnamefont {M.}~\bibnamefont {Rini}}, \bibinfo {author} {\bibfnamefont
  {T.~E.}\ \bibnamefont {Glover}}, \bibinfo {author} {\bibfnamefont {M.~P.}\
  \bibnamefont {Hertlein}}, \bibinfo {author} {\bibfnamefont {A.~G.~C.}\
  \bibnamefont {Gonzalez}}, \bibinfo {author} {\bibfnamefont {N.}~\bibnamefont
  {Tahir}}, \bibinfo {author} {\bibfnamefont {Y.}~\bibnamefont {Tomioka}},
  \bibinfo {author} {\bibfnamefont {Y.}~\bibnamefont {Tokura}}, \bibinfo
  {author} {\bibfnamefont {Z.}~\bibnamefont {Hussain}}, \ and\ \bibinfo
  {author} {\bibfnamefont {R.~W.}\ \bibnamefont {Schoenlein}},\ }\href
  {https://doi.org/10.1038/srep04050} {\bibfield  {journal} {\bibinfo
  {journal} {Scientific Reports}\ }\textbf {\bibinfo {volume} {4}},\ \bibinfo
  {pages} {4050} (\bibinfo {year} {2014})}\BibitemShut {NoStop}%
\bibitem [{\citenamefont {Fujioka}\ \emph {et~al.}(2010)\citenamefont
  {Fujioka}, \citenamefont {Ida}, \citenamefont {Takahashi}, \citenamefont
  {Kida}, \citenamefont {Shimano},\ and\ \citenamefont {Tokura}}]{Fujioka}%
  \BibitemOpen
  \bibfield  {author} {\bibinfo {author} {\bibfnamefont {J.}~\bibnamefont
  {Fujioka}}, \bibinfo {author} {\bibfnamefont {Y.}~\bibnamefont {Ida}},
  \bibinfo {author} {\bibfnamefont {Y.}~\bibnamefont {Takahashi}}, \bibinfo
  {author} {\bibfnamefont {N.}~\bibnamefont {Kida}}, \bibinfo {author}
  {\bibfnamefont {R.}~\bibnamefont {Shimano}}, \ and\ \bibinfo {author}
  {\bibfnamefont {Y.}~\bibnamefont {Tokura}},\ }\href {\doibase
  10.1103/PhysRevB.82.140409} {\bibfield  {journal} {\bibinfo  {journal} {Phys.
  Rev. B}\ }\textbf {\bibinfo {volume} {82}},\ \bibinfo {pages} {140409}
  (\bibinfo {year} {2010})}\BibitemShut {NoStop}%
\bibitem [{\citenamefont {Zaanen}\ \emph {et~al.}(1985)\citenamefont {Zaanen},
  \citenamefont {Sawatzky},\ and\ \citenamefont {Allen}}]{Jan}%
  \BibitemOpen
  \bibfield  {author} {\bibinfo {author} {\bibfnamefont {J.}~\bibnamefont
  {Zaanen}}, \bibinfo {author} {\bibfnamefont {G.~A.}\ \bibnamefont
  {Sawatzky}}, \ and\ \bibinfo {author} {\bibfnamefont {J.~W.}\ \bibnamefont
  {Allen}},\ }\href {\doibase 10.1103/PhysRevLett.55.418} {\bibfield  {journal}
  {\bibinfo  {journal} {Phys. Rev. Lett.}\ }\textbf {\bibinfo {volume} {55}},\
  \bibinfo {pages} {418} (\bibinfo {year} {1985})}\BibitemShut {NoStop}%
\bibitem [{\citenamefont {Arima}\ \emph {et~al.}(1993)\citenamefont {Arima},
  \citenamefont {Tokura},\ and\ \citenamefont {Torrance}}]{Arima}%
  \BibitemOpen
  \bibfield  {author} {\bibinfo {author} {\bibfnamefont {T.}~\bibnamefont
  {Arima}}, \bibinfo {author} {\bibfnamefont {Y.}~\bibnamefont {Tokura}}, \
  and\ \bibinfo {author} {\bibfnamefont {J.~B.}\ \bibnamefont {Torrance}},\
  }\href {\doibase 10.1103/PhysRevB.48.17006} {\bibfield  {journal} {\bibinfo
  {journal} {Phys. Rev. B}\ }\textbf {\bibinfo {volume} {48}},\ \bibinfo
  {pages} {17006} (\bibinfo {year} {1993})}\BibitemShut {NoStop}%
\bibitem [{\citenamefont {G\"ossling}\ \emph {et~al.}(2008)\citenamefont
  {G\"ossling}, \citenamefont {Haverkort}, \citenamefont {Benomar},
  \citenamefont {Wu}, \citenamefont {Senff}, \citenamefont {M\"oller},
  \citenamefont {Braden}, \citenamefont {Mydosh},\ and\ \citenamefont
  {Gr\"uninger}}]{Gossling}%
  \BibitemOpen
  \bibfield  {author} {\bibinfo {author} {\bibfnamefont {A.}~\bibnamefont
  {G\"ossling}}, \bibinfo {author} {\bibfnamefont {M.~W.}\ \bibnamefont
  {Haverkort}}, \bibinfo {author} {\bibfnamefont {M.}~\bibnamefont {Benomar}},
  \bibinfo {author} {\bibfnamefont {H.}~\bibnamefont {Wu}}, \bibinfo {author}
  {\bibfnamefont {D.}~\bibnamefont {Senff}}, \bibinfo {author} {\bibfnamefont
  {T.}~\bibnamefont {M\"oller}}, \bibinfo {author} {\bibfnamefont
  {M.}~\bibnamefont {Braden}}, \bibinfo {author} {\bibfnamefont {J.~A.}\
  \bibnamefont {Mydosh}}, \ and\ \bibinfo {author} {\bibfnamefont
  {M.}~\bibnamefont {Gr\"uninger}},\ }\href {\doibase
  10.1103/PhysRevB.77.035109} {\bibfield  {journal} {\bibinfo  {journal} {Phys.
  Rev. B}\ }\textbf {\bibinfo {volume} {77}},\ \bibinfo {pages} {035109}
  (\bibinfo {year} {2008})}\BibitemShut {NoStop}%
\bibitem [{\citenamefont {Mildner}\ \emph {et~al.}(2015)\citenamefont
  {Mildner}, \citenamefont {Hoffmann}, \citenamefont {Bl\"ochl}, \citenamefont
  {Techert},\ and\ \citenamefont {Jooss}}]{Mildner}%
  \BibitemOpen
  \bibfield  {author} {\bibinfo {author} {\bibfnamefont {S.}~\bibnamefont
  {Mildner}}, \bibinfo {author} {\bibfnamefont {J.}~\bibnamefont {Hoffmann}},
  \bibinfo {author} {\bibfnamefont {P.~E.}\ \bibnamefont {Bl\"ochl}}, \bibinfo
  {author} {\bibfnamefont {S.}~\bibnamefont {Techert}}, \ and\ \bibinfo
  {author} {\bibfnamefont {C.}~\bibnamefont {Jooss}},\ }\href {\doibase
  10.1103/PhysRevB.92.035145} {\bibfield  {journal} {\bibinfo  {journal} {Phys.
  Rev. B}\ }\textbf {\bibinfo {volume} {92}},\ \bibinfo {pages} {035145}
  (\bibinfo {year} {2015})}\BibitemShut {NoStop}%
\bibitem [{JTN()}]{JTNote}%
  \BibitemOpen
  \href@noop {} {}\bibinfo {note} {In the case of a transient JT distortion on
  the $e_\mathrm{g}$ orbitals that are degenerate in equilibrium, still no
  additional long-range orbital order can be probed on the femto- to pico-
  second timescale because the formation of its spatial coherence is limited by
  the speed of sound \cite{Beaud}.}\BibitemShut {Stop}%
\bibitem [{\citenamefont {Wilkins}\ \emph {et~al.}(2003)\citenamefont
  {Wilkins}, \citenamefont {Spencer}, \citenamefont {Hatton}, \citenamefont
  {Collins}, \citenamefont {Roper}, \citenamefont {Prabhakaran},\ and\
  \citenamefont {Boothroyd}}]{wilkins-2003-prl}%
  \BibitemOpen
  \bibfield  {author} {\bibinfo {author} {\bibfnamefont {S.~B.}\ \bibnamefont
  {Wilkins}}, \bibinfo {author} {\bibfnamefont {P.~D.}\ \bibnamefont
  {Spencer}}, \bibinfo {author} {\bibfnamefont {P.~D.}\ \bibnamefont {Hatton}},
  \bibinfo {author} {\bibfnamefont {S.~P.}\ \bibnamefont {Collins}}, \bibinfo
  {author} {\bibfnamefont {M.~D.}\ \bibnamefont {Roper}}, \bibinfo {author}
  {\bibfnamefont {D.}~\bibnamefont {Prabhakaran}}, \ and\ \bibinfo {author}
  {\bibfnamefont {A.~T.}\ \bibnamefont {Boothroyd}},\ }\href {\doibase
  10.1103/PhysRevLett.91.167205} {\bibfield  {journal} {\bibinfo  {journal}
  {Phys. Rev. Lett.}\ }\textbf {\bibinfo {volume} {91}},\ \bibinfo {pages}
  {167205} (\bibinfo {year} {2003})}\BibitemShut {NoStop}%
\bibitem [{\citenamefont {Kimura}\ \emph {et~al.}(2001)\citenamefont {Kimura},
  \citenamefont {Hatsuda}, \citenamefont {Ueno}, \citenamefont {Kajimoto},
  \citenamefont {Mochizuki}, \citenamefont {Yoshizawa}, \citenamefont {Nagai},
  \citenamefont {Matsui}, \citenamefont {Yamazaki},\ and\ \citenamefont
  {Tokura}}]{Kimura}%
  \BibitemOpen
  \bibfield  {author} {\bibinfo {author} {\bibfnamefont {T.}~\bibnamefont
  {Kimura}}, \bibinfo {author} {\bibfnamefont {K.}~\bibnamefont {Hatsuda}},
  \bibinfo {author} {\bibfnamefont {Y.}~\bibnamefont {Ueno}}, \bibinfo {author}
  {\bibfnamefont {R.}~\bibnamefont {Kajimoto}}, \bibinfo {author}
  {\bibfnamefont {H.}~\bibnamefont {Mochizuki}}, \bibinfo {author}
  {\bibfnamefont {H.}~\bibnamefont {Yoshizawa}}, \bibinfo {author}
  {\bibfnamefont {T.}~\bibnamefont {Nagai}}, \bibinfo {author} {\bibfnamefont
  {Y.}~\bibnamefont {Matsui}}, \bibinfo {author} {\bibfnamefont
  {A.}~\bibnamefont {Yamazaki}}, \ and\ \bibinfo {author} {\bibfnamefont
  {Y.}~\bibnamefont {Tokura}},\ }\href {\doibase 10.1103/PhysRevB.65.020407}
  {\bibfield  {journal} {\bibinfo  {journal} {Phys. Rev. B}\ }\textbf {\bibinfo
  {volume} {65}},\ \bibinfo {pages} {020407} (\bibinfo {year}
  {2001})}\BibitemShut {NoStop}%
\bibitem [{\citenamefont {Dakovski}\ \emph {et~al.}(2015)\citenamefont
  {Dakovski}, \citenamefont {Heimann}, \citenamefont {Holmes}, \citenamefont
  {Krupin}, \citenamefont {Minitti}, \citenamefont {Mitra}, \citenamefont
  {Moeller}, \citenamefont {Rowen}, \citenamefont {Schlotter},\ and\
  \citenamefont {Turner}}]{Dakovski-2015-JSR}%
  \BibitemOpen
  \bibfield  {author} {\bibinfo {author} {\bibfnamefont {G.~L.}\ \bibnamefont
  {Dakovski}}, \bibinfo {author} {\bibfnamefont {P.}~\bibnamefont {Heimann}},
  \bibinfo {author} {\bibfnamefont {M.}~\bibnamefont {Holmes}}, \bibinfo
  {author} {\bibfnamefont {O.}~\bibnamefont {Krupin}}, \bibinfo {author}
  {\bibfnamefont {M.~P.}\ \bibnamefont {Minitti}}, \bibinfo {author}
  {\bibfnamefont {A.}~\bibnamefont {Mitra}}, \bibinfo {author} {\bibfnamefont
  {S.}~\bibnamefont {Moeller}}, \bibinfo {author} {\bibfnamefont
  {M.}~\bibnamefont {Rowen}}, \bibinfo {author} {\bibfnamefont {W.~F.}\
  \bibnamefont {Schlotter}}, \ and\ \bibinfo {author} {\bibfnamefont {J.~J.}\
  \bibnamefont {Turner}},\ }\href@noop {} {\bibfield  {journal} {\bibinfo
  {journal} {J. Synchrotron Rad.}\ }\textbf {\bibinfo {volume} {22}},\ \bibinfo
  {pages} {498} (\bibinfo {year} {2015})}\BibitemShut {NoStop}%
\bibitem [{\citenamefont {Bostedt}\ \emph {et~al.}(2016)\citenamefont
  {Bostedt}, \citenamefont {Boutet}, \citenamefont {Fritz}, \citenamefont
  {Huang}, \citenamefont {Lee}, \citenamefont {Lemke}, \citenamefont {Robert},
  \citenamefont {Schlotter}, \citenamefont {Turner},\ and\ \citenamefont
  {Williams}}]{Bostedt-2016-RMP}%
  \BibitemOpen
  \bibfield  {author} {\bibinfo {author} {\bibfnamefont {C.}~\bibnamefont
  {Bostedt}}, \bibinfo {author} {\bibfnamefont {S.}~\bibnamefont {Boutet}},
  \bibinfo {author} {\bibfnamefont {D.~M.}\ \bibnamefont {Fritz}}, \bibinfo
  {author} {\bibfnamefont {Z.}~\bibnamefont {Huang}}, \bibinfo {author}
  {\bibfnamefont {H.~J.}\ \bibnamefont {Lee}}, \bibinfo {author} {\bibfnamefont
  {H.~T.}\ \bibnamefont {Lemke}}, \bibinfo {author} {\bibfnamefont
  {A.}~\bibnamefont {Robert}}, \bibinfo {author} {\bibfnamefont {W.~F.}\
  \bibnamefont {Schlotter}}, \bibinfo {author} {\bibfnamefont {J.~J.}\
  \bibnamefont {Turner}}, \ and\ \bibinfo {author} {\bibfnamefont {G.~J.}\
  \bibnamefont {Williams}},\ }\href {\doibase 10.1103/RevModPhys.88.015007}
  {\bibfield  {journal} {\bibinfo  {journal} {Rev. Mod. Phys.}\ }\textbf
  {\bibinfo {volume} {88}},\ \bibinfo {pages} {015007} (\bibinfo {year}
  {2016})}\BibitemShut {NoStop}%
\bibitem [{\citenamefont {Doering}\ \emph {et~al.}(2011)\citenamefont
  {Doering}, \citenamefont {Chuang}, \citenamefont {Andresen}, \citenamefont
  {Chow}, \citenamefont {Contarato}, \citenamefont {Cummings}, \citenamefont
  {Domning}, \citenamefont {Joseph}, \citenamefont {Pepper}, \citenamefont
  {Smith} \emph {et~al.}}]{Doering-2011-RSI}%
  \BibitemOpen
  \bibfield  {author} {\bibinfo {author} {\bibfnamefont {D.}~\bibnamefont
  {Doering}}, \bibinfo {author} {\bibfnamefont {Y.~D.}\ \bibnamefont {Chuang}},
  \bibinfo {author} {\bibfnamefont {N.}~\bibnamefont {Andresen}}, \bibinfo
  {author} {\bibfnamefont {K.}~\bibnamefont {Chow}}, \bibinfo {author}
  {\bibfnamefont {D.}~\bibnamefont {Contarato}}, \bibinfo {author}
  {\bibfnamefont {C.}~\bibnamefont {Cummings}}, \bibinfo {author}
  {\bibfnamefont {E.}~\bibnamefont {Domning}}, \bibinfo {author} {\bibfnamefont
  {J.}~\bibnamefont {Joseph}}, \bibinfo {author} {\bibfnamefont {J.~S.}\
  \bibnamefont {Pepper}}, \bibinfo {author} {\bibfnamefont {B.}~\bibnamefont
  {Smith}},  \emph {et~al.},\ }\href@noop {} {\bibfield  {journal} {\bibinfo
  {journal} {Rev. Sci. Instrum.}\ }\textbf {\bibinfo {volume} {82}},\ \bibinfo
  {pages} {073303} (\bibinfo {year} {2011})}\BibitemShut {NoStop}%
\bibitem [{\citenamefont {Turner}\ \emph {et~al.}(2015)\citenamefont {Turner},
  \citenamefont {Dakovski}, \citenamefont {Hoffmann}, \citenamefont {Hwang},
  \citenamefont {Zarem}, \citenamefont {Schlotter}, \citenamefont {Moeller},
  \citenamefont {Minitti}, \citenamefont {Staub}, \citenamefont {Johnson} \emph
  {et~al.}}]{Turner-2015-JSR}%
  \BibitemOpen
  \bibfield  {author} {\bibinfo {author} {\bibfnamefont {J.~J.}\ \bibnamefont
  {Turner}}, \bibinfo {author} {\bibfnamefont {G.~L.}\ \bibnamefont
  {Dakovski}}, \bibinfo {author} {\bibfnamefont {M.}~\bibnamefont {Hoffmann}},
  \bibinfo {author} {\bibfnamefont {H.~Y.}\ \bibnamefont {Hwang}}, \bibinfo
  {author} {\bibfnamefont {A.}~\bibnamefont {Zarem}}, \bibinfo {author}
  {\bibfnamefont {W.~F.}\ \bibnamefont {Schlotter}}, \bibinfo {author}
  {\bibfnamefont {S.}~\bibnamefont {Moeller}}, \bibinfo {author} {\bibfnamefont
  {M.~P.}\ \bibnamefont {Minitti}}, \bibinfo {author} {\bibfnamefont
  {U.}~\bibnamefont {Staub}}, \bibinfo {author} {\bibfnamefont
  {S.}~\bibnamefont {Johnson}},  \emph {et~al.},\ }\href@noop {} {\bibfield
  {journal} {\bibinfo  {journal} {J. Synchrotron Rad.}\ }\textbf {\bibinfo
  {volume} {22}},\ \bibinfo {pages} {621} (\bibinfo {year} {2015})}\BibitemShut
  {NoStop}%
\bibitem [{\citenamefont {Tiedtke}\ \emph {et~al.}(2014)\citenamefont
  {Tiedtke}, \citenamefont {Sorokin}, \citenamefont {Jastrow}, \citenamefont
  {Juranic}, \citenamefont {Kreis}, \citenamefont {Gerken}, \citenamefont
  {Richter}, \citenamefont {Arp}, \citenamefont {Feng},\ and\ \citenamefont
  {Nordlund}}]{Tiedtke-2014-OptExpress}%
  \BibitemOpen
  \bibfield  {author} {\bibinfo {author} {\bibfnamefont {K.}~\bibnamefont
  {Tiedtke}}, \bibinfo {author} {\bibfnamefont {A.~A.}\ \bibnamefont
  {Sorokin}}, \bibinfo {author} {\bibfnamefont {U.}~\bibnamefont {Jastrow}},
  \bibinfo {author} {\bibfnamefont {P.}~\bibnamefont {Juranic}}, \bibinfo
  {author} {\bibfnamefont {S.}~\bibnamefont {Kreis}}, \bibinfo {author}
  {\bibfnamefont {N.}~\bibnamefont {Gerken}}, \bibinfo {author} {\bibfnamefont
  {M.}~\bibnamefont {Richter}}, \bibinfo {author} {\bibfnamefont
  {U.}~\bibnamefont {Arp}}, \bibinfo {author} {\bibfnamefont {Y.}~\bibnamefont
  {Feng}}, \ and\ \bibinfo {author} {\bibfnamefont {D.}~\bibnamefont
  {Nordlund}},\ }\href
  {http://www.opticsexpress.org/abstract.cfm?URI=oe-22-18-21214} {\bibfield
  {journal} {\bibinfo  {journal} {Opt. Express}\ }\textbf {\bibinfo {volume}
  {22}},\ \bibinfo {pages} {21214} (\bibinfo {year} {2014})}\BibitemShut
  {NoStop}%
\bibitem [{\citenamefont {Krupin}\ \emph {et~al.}(2012)\citenamefont {Krupin},
  \citenamefont {Trigo}, \citenamefont {Schlotter}, \citenamefont {Beye},
  \citenamefont {Sorgenfrei}, \citenamefont {Turner}, \citenamefont {Reis},
  \citenamefont {Gerken}, \citenamefont {Lee}, \citenamefont {Lee} \emph
  {et~al.}}]{Krupin-2012-OptExp}%
  \BibitemOpen
  \bibfield  {author} {\bibinfo {author} {\bibfnamefont {O.}~\bibnamefont
  {Krupin}}, \bibinfo {author} {\bibfnamefont {M.}~\bibnamefont {Trigo}},
  \bibinfo {author} {\bibfnamefont {W.~F.}\ \bibnamefont {Schlotter}}, \bibinfo
  {author} {\bibfnamefont {M.}~\bibnamefont {Beye}}, \bibinfo {author}
  {\bibfnamefont {F.}~\bibnamefont {Sorgenfrei}}, \bibinfo {author}
  {\bibfnamefont {J.~J.}\ \bibnamefont {Turner}}, \bibinfo {author}
  {\bibfnamefont {D.~A.}\ \bibnamefont {Reis}}, \bibinfo {author}
  {\bibfnamefont {N.}~\bibnamefont {Gerken}}, \bibinfo {author} {\bibfnamefont
  {S.}~\bibnamefont {Lee}}, \bibinfo {author} {\bibfnamefont {W.~S.}\
  \bibnamefont {Lee}},  \emph {et~al.},\ }\href@noop {} {\bibfield  {journal}
  {\bibinfo  {journal} {Opt. Express}\ }\textbf {\bibinfo {volume} {20}},\
  \bibinfo {pages} {11396} (\bibinfo {year} {2012})}\BibitemShut {NoStop}%
\bibitem [{Note1()}]{Note1}%
  \BibitemOpen
  \bibinfo {note} {We have carefully calculated the refractive index at the NIR
  pump energy based on the optical conductivity (Fig. \ref {fig:1}c) to obtain
  a value of 2.62, giving a penetration depth of 67\protect \tmspace
  +\thinmuskip {.1667em}nm. This value was then used for the energy density
  calculation. The incident pump angles in different geometries have also been
  taken into account, producing internal angles leading only to small
  corrections of the energy density value ($<$ 10 $\%$), within the energy
  density errors.}\BibitemShut {Stop}%
\bibitem [{\citenamefont {Ehrke}\ \emph {et~al.}(2011)\citenamefont {Ehrke},
  \citenamefont {Tobey}, \citenamefont {Wall}, \citenamefont {Cavill},
  \citenamefont {F\"orst}, \citenamefont {Khanna}, \citenamefont {Garl},
  \citenamefont {Stojanovic}, \citenamefont {Prabhakaran}, \citenamefont
  {Boothroyd}, \citenamefont {Gensch}, \citenamefont {Mirone}, \citenamefont
  {Reutler}, \citenamefont {Revcolevschi}, \citenamefont {Dhesi},\ and\
  \citenamefont {Cavalleri}}]{Ehrke}%
  \BibitemOpen
  \bibfield  {author} {\bibinfo {author} {\bibfnamefont {H.}~\bibnamefont
  {Ehrke}}, \bibinfo {author} {\bibfnamefont {R.~I.}\ \bibnamefont {Tobey}},
  \bibinfo {author} {\bibfnamefont {S.}~\bibnamefont {Wall}}, \bibinfo {author}
  {\bibfnamefont {S.~A.}\ \bibnamefont {Cavill}}, \bibinfo {author}
  {\bibfnamefont {M.}~\bibnamefont {F\"orst}}, \bibinfo {author} {\bibfnamefont
  {V.}~\bibnamefont {Khanna}}, \bibinfo {author} {\bibfnamefont
  {T.}~\bibnamefont {Garl}}, \bibinfo {author} {\bibfnamefont {N.}~\bibnamefont
  {Stojanovic}}, \bibinfo {author} {\bibfnamefont {D.}~\bibnamefont
  {Prabhakaran}}, \bibinfo {author} {\bibfnamefont {A.~T.}\ \bibnamefont
  {Boothroyd}}, \bibinfo {author} {\bibfnamefont {M.}~\bibnamefont {Gensch}},
  \bibinfo {author} {\bibfnamefont {A.}~\bibnamefont {Mirone}}, \bibinfo
  {author} {\bibfnamefont {P.}~\bibnamefont {Reutler}}, \bibinfo {author}
  {\bibfnamefont {A.}~\bibnamefont {Revcolevschi}}, \bibinfo {author}
  {\bibfnamefont {S.~S.}\ \bibnamefont {Dhesi}}, \ and\ \bibinfo {author}
  {\bibfnamefont {A.}~\bibnamefont {Cavalleri}},\ }\href {\doibase
  10.1103/PhysRevLett.106.217401} {\bibfield  {journal} {\bibinfo  {journal}
  {Phys. Rev. Lett.}\ }\textbf {\bibinfo {volume} {106}},\ \bibinfo {pages}
  {217401} (\bibinfo {year} {2011})}\BibitemShut {NoStop}%
\bibitem [{\citenamefont {Kovaleva}\ \emph {et~al.}(2004)\citenamefont
  {Kovaleva}, \citenamefont {Boris}, \citenamefont {Bernhard}, \citenamefont
  {Kulakov}, \citenamefont {Pimenov}, \citenamefont {Balbashov}, \citenamefont
  {Khaliullin},\ and\ \citenamefont {Keimer}}]{Kovaleva}%
  \BibitemOpen
  \bibfield  {author} {\bibinfo {author} {\bibfnamefont {N.~N.}\ \bibnamefont
  {Kovaleva}}, \bibinfo {author} {\bibfnamefont {A.~V.}\ \bibnamefont {Boris}},
  \bibinfo {author} {\bibfnamefont {C.}~\bibnamefont {Bernhard}}, \bibinfo
  {author} {\bibfnamefont {A.}~\bibnamefont {Kulakov}}, \bibinfo {author}
  {\bibfnamefont {A.}~\bibnamefont {Pimenov}}, \bibinfo {author} {\bibfnamefont
  {A.~M.}\ \bibnamefont {Balbashov}}, \bibinfo {author} {\bibfnamefont
  {G.}~\bibnamefont {Khaliullin}}, \ and\ \bibinfo {author} {\bibfnamefont
  {B.}~\bibnamefont {Keimer}},\ }\href {\doibase 10.1103/PhysRevLett.93.147204}
  {\bibfield  {journal} {\bibinfo  {journal} {Phys. Rev. Lett.}\ }\textbf
  {\bibinfo {volume} {93}},\ \bibinfo {pages} {147204} (\bibinfo {year}
  {2004})}\BibitemShut {NoStop}%
\bibitem [{\citenamefont {Tobe}\ \emph {et~al.}(2001)\citenamefont {Tobe},
  \citenamefont {Kimura}, \citenamefont {Okimoto},\ and\ \citenamefont
  {Tokura}}]{Tobe}%
  \BibitemOpen
  \bibfield  {author} {\bibinfo {author} {\bibfnamefont {K.}~\bibnamefont
  {Tobe}}, \bibinfo {author} {\bibfnamefont {T.}~\bibnamefont {Kimura}},
  \bibinfo {author} {\bibfnamefont {Y.}~\bibnamefont {Okimoto}}, \ and\
  \bibinfo {author} {\bibfnamefont {Y.}~\bibnamefont {Tokura}},\ }\href
  {\doibase 10.1103/PhysRevB.64.184421} {\bibfield  {journal} {\bibinfo
  {journal} {Phys. Rev. B}\ }\textbf {\bibinfo {volume} {64}},\ \bibinfo
  {pages} {184421} (\bibinfo {year} {2001})}\BibitemShut {NoStop}%
\bibitem [{\citenamefont {Averitt}\ and\ \citenamefont
  {Taylor}(2002)}]{Averitt}%
  \BibitemOpen
  \bibfield  {author} {\bibinfo {author} {\bibfnamefont {R.~D.}\ \bibnamefont
  {Averitt}}\ and\ \bibinfo {author} {\bibfnamefont {A.~J.}\ \bibnamefont
  {Taylor}},\ }\href {\doibase 10.1088/0953-8984/14/50/203} {\bibfield
  {journal} {\bibinfo  {journal} {Journal of Physics: Condensed Matter}\
  }\textbf {\bibinfo {volume} {14}},\ \bibinfo {pages} {R1357} (\bibinfo {year}
  {2002})}\BibitemShut {NoStop}%
\bibitem [{\citenamefont {Esposito}\ \emph {et~al.}(2018)\citenamefont
  {Esposito}, \citenamefont {Rettig}, \citenamefont {Abreu}, \citenamefont
  {Bothschafter}, \citenamefont {Ingold}, \citenamefont {Kawasaki},
  \citenamefont {Kubli}, \citenamefont {Lantz}, \citenamefont {Nakamura},
  \citenamefont {Rittman}, \citenamefont {Savoini}, \citenamefont {Tokura},
  \citenamefont {Staub}, \citenamefont {Johnson},\ and\ \citenamefont
  {Beaud}}]{Esposito}%
  \BibitemOpen
  \bibfield  {author} {\bibinfo {author} {\bibfnamefont {V.}~\bibnamefont
  {Esposito}}, \bibinfo {author} {\bibfnamefont {L.}~\bibnamefont {Rettig}},
  \bibinfo {author} {\bibfnamefont {E.}~\bibnamefont {Abreu}}, \bibinfo
  {author} {\bibfnamefont {E.~M.}\ \bibnamefont {Bothschafter}}, \bibinfo
  {author} {\bibfnamefont {G.}~\bibnamefont {Ingold}}, \bibinfo {author}
  {\bibfnamefont {M.}~\bibnamefont {Kawasaki}}, \bibinfo {author}
  {\bibfnamefont {M.}~\bibnamefont {Kubli}}, \bibinfo {author} {\bibfnamefont
  {G.}~\bibnamefont {Lantz}}, \bibinfo {author} {\bibfnamefont
  {M.}~\bibnamefont {Nakamura}}, \bibinfo {author} {\bibfnamefont
  {J.}~\bibnamefont {Rittman}}, \bibinfo {author} {\bibfnamefont
  {M.}~\bibnamefont {Savoini}}, \bibinfo {author} {\bibfnamefont
  {Y.}~\bibnamefont {Tokura}}, \bibinfo {author} {\bibfnamefont
  {U.}~\bibnamefont {Staub}}, \bibinfo {author} {\bibfnamefont {S.~L.}\
  \bibnamefont {Johnson}}, \ and\ \bibinfo {author} {\bibfnamefont
  {P.}~\bibnamefont {Beaud}},\ }\href {\doibase 10.1103/PhysRevB.97.014312}
  {\bibfield  {journal} {\bibinfo  {journal} {Phys. Rev. B}\ }\textbf {\bibinfo
  {volume} {97}},\ \bibinfo {pages} {014312} (\bibinfo {year}
  {2018})}\BibitemShut {NoStop}%
\bibitem [{\citenamefont {Beye}\ \emph {et~al.}(2012)\citenamefont {Beye},
  \citenamefont {Krupin}, \citenamefont {Hays}, \citenamefont {Reid},
  \citenamefont {Rupp}, \citenamefont {Jong}, \citenamefont {Lee},
  \citenamefont {Lee}, \citenamefont {Chuang}, \citenamefont {Coffee} \emph
  {et~al.}}]{Beye-2012-APL}%
  \BibitemOpen
  \bibfield  {author} {\bibinfo {author} {\bibfnamefont {M.}~\bibnamefont
  {Beye}}, \bibinfo {author} {\bibfnamefont {O.}~\bibnamefont {Krupin}},
  \bibinfo {author} {\bibfnamefont {G.}~\bibnamefont {Hays}}, \bibinfo {author}
  {\bibfnamefont {A.~H.}\ \bibnamefont {Reid}}, \bibinfo {author}
  {\bibfnamefont {D.}~\bibnamefont {Rupp}}, \bibinfo {author} {\bibfnamefont
  {S.~d.}\ \bibnamefont {Jong}}, \bibinfo {author} {\bibfnamefont
  {S.}~\bibnamefont {Lee}}, \bibinfo {author} {\bibfnamefont {W.~S.}\
  \bibnamefont {Lee}}, \bibinfo {author} {\bibfnamefont {Y.~D.}\ \bibnamefont
  {Chuang}}, \bibinfo {author} {\bibfnamefont {R.}~\bibnamefont {Coffee}},
  \emph {et~al.},\ }\href {\doibase 10.1063/1.3695164} {\bibfield  {journal}
  {\bibinfo  {journal} {Appl. Phys. Lett.}\ }\textbf {\bibinfo {volume}
  {100}},\ \bibinfo {pages} {121108} (\bibinfo {year} {2012})}\BibitemShut
  {NoStop}%
\bibitem [{\citenamefont {Ulbrich}\ \emph {et~al.}(2012)\citenamefont
  {Ulbrich}, \citenamefont {Steffens}, \citenamefont {Lamago}, \citenamefont
  {Sidis},\ and\ \citenamefont {Braden}}]{Ulbrich2}%
  \BibitemOpen
  \bibfield  {author} {\bibinfo {author} {\bibfnamefont {H.}~\bibnamefont
  {Ulbrich}}, \bibinfo {author} {\bibfnamefont {P.}~\bibnamefont {Steffens}},
  \bibinfo {author} {\bibfnamefont {D.}~\bibnamefont {Lamago}}, \bibinfo
  {author} {\bibfnamefont {Y.}~\bibnamefont {Sidis}}, \ and\ \bibinfo {author}
  {\bibfnamefont {M.}~\bibnamefont {Braden}},\ }\href {\doibase
  10.1103/PhysRevLett.108.247209} {\bibfield  {journal} {\bibinfo  {journal}
  {Phys. Rev. Lett.}\ }\textbf {\bibinfo {volume} {108}},\ \bibinfo {pages}
  {247209} (\bibinfo {year} {2012})}\BibitemShut {NoStop}%
\bibitem [{\citenamefont {Ulbrich}\ \emph {et~al.}(2011)\citenamefont
  {Ulbrich}, \citenamefont {Senff}, \citenamefont {Steffens}, \citenamefont
  {Schumann}, \citenamefont {Sidis}, \citenamefont {Reutler}, \citenamefont
  {Revcolevschi},\ and\ \citenamefont {Braden}}]{Ulbrich}%
  \BibitemOpen
  \bibfield  {author} {\bibinfo {author} {\bibfnamefont {H.}~\bibnamefont
  {Ulbrich}}, \bibinfo {author} {\bibfnamefont {D.}~\bibnamefont {Senff}},
  \bibinfo {author} {\bibfnamefont {P.}~\bibnamefont {Steffens}}, \bibinfo
  {author} {\bibfnamefont {O.~J.}\ \bibnamefont {Schumann}}, \bibinfo {author}
  {\bibfnamefont {Y.}~\bibnamefont {Sidis}}, \bibinfo {author} {\bibfnamefont
  {P.}~\bibnamefont {Reutler}}, \bibinfo {author} {\bibfnamefont
  {A.}~\bibnamefont {Revcolevschi}}, \ and\ \bibinfo {author} {\bibfnamefont
  {M.}~\bibnamefont {Braden}},\ }\href {\doibase
  10.1103/PhysRevLett.106.157201} {\bibfield  {journal} {\bibinfo  {journal}
  {Phys. Rev. Lett.}\ }\textbf {\bibinfo {volume} {106}},\ \bibinfo {pages}
  {157201} (\bibinfo {year} {2011})}\BibitemShut {NoStop}%
\bibitem [{\citenamefont {Spaldin}\ and\ \citenamefont
  {Ramesh}(2019)}]{Spaldin}%
  \BibitemOpen
  \bibfield  {author} {\bibinfo {author} {\bibfnamefont {N.~A.}\ \bibnamefont
  {Spaldin}}\ and\ \bibinfo {author} {\bibfnamefont {R.}~\bibnamefont
  {Ramesh}},\ }\href {\doibase 10.1038/s41563-018-0275-2} {\bibfield  {journal}
  {\bibinfo  {journal} {Nature Materials}\ }\textbf {\bibinfo {volume} {18}},\
  \bibinfo {pages} {203} (\bibinfo {year} {2019})}\BibitemShut {NoStop}%
\bibitem [{\citenamefont {Giannetti}\ \emph {et~al.}(2011)\citenamefont
  {Giannetti}, \citenamefont {Cilento}, \citenamefont {Conte}, \citenamefont
  {Coslovich}, \citenamefont {Ferrini}, \citenamefont {Molegraaf},
  \citenamefont {Raichle}, \citenamefont {Liang}, \citenamefont {Eisaki},
  \citenamefont {Greven}, \citenamefont {Damascelli}, \citenamefont {van~der
  Marel},\ and\ \citenamefont {Parmigiani}}]{giannetti-natcomm-2011}%
  \BibitemOpen
  \bibfield  {author} {\bibinfo {author} {\bibfnamefont {C.}~\bibnamefont
  {Giannetti}}, \bibinfo {author} {\bibfnamefont {F.}~\bibnamefont {Cilento}},
  \bibinfo {author} {\bibfnamefont {S.~D.}\ \bibnamefont {Conte}}, \bibinfo
  {author} {\bibfnamefont {G.}~\bibnamefont {Coslovich}}, \bibinfo {author}
  {\bibfnamefont {G.}~\bibnamefont {Ferrini}}, \bibinfo {author} {\bibfnamefont
  {H.}~\bibnamefont {Molegraaf}}, \bibinfo {author} {\bibfnamefont
  {M.}~\bibnamefont {Raichle}}, \bibinfo {author} {\bibfnamefont
  {R.}~\bibnamefont {Liang}}, \bibinfo {author} {\bibfnamefont
  {H.}~\bibnamefont {Eisaki}}, \bibinfo {author} {\bibfnamefont
  {M.}~\bibnamefont {Greven}}, \bibinfo {author} {\bibfnamefont
  {A.}~\bibnamefont {Damascelli}}, \bibinfo {author} {\bibfnamefont
  {D.}~\bibnamefont {van~der Marel}}, \ and\ \bibinfo {author} {\bibfnamefont
  {F.}~\bibnamefont {Parmigiani}},\ }\href {\doibase 10.1038/ncomms1354}
  {\bibfield  {journal} {\bibinfo  {journal} {Nature Communications}\ }\textbf
  {\bibinfo {volume} {2}},\ \bibinfo {pages} {353} (\bibinfo {year}
  {2011})}\BibitemShut {NoStop}%
\end{thebibliography}

\end{document}